%% file: main.tex
\documentclass[
a4paper,
amsfonts,
amsmath,
reprint,
showkeys,
nofootinbib,
twoside,
onecolumn,
notitlepage
]{revtex4-1}
\usepackage{amsmath,amstext}
\usepackage[T1]{fontenc}
\usepackage{amssymb}
\input{epsf}
\usepackage{graphicx}
\usepackage{ae,aecompl}

\usepackage{hyperref}
\usepackage{amsmath}
\usepackage{amssymb}
\usepackage{mathtools}
\usepackage{bm}
\usepackage{cleveref}
\usepackage{tensor}
\usepackage{braket}
\usepackage{enumitem}
\usepackage{mhchem}
\usepackage{cancel}
\usepackage{physics}
\usepackage{color}

\newcommand{\spc}{\quad \quad \quad}

\newcommand{\n}{{\bm{n}}}
\newcommand{\m}{{\bm{\mu}}}

\input{commands}

\begin{document}
	
	\title{Stability of multicomponent Israel-Stewart-Maxwell theory for charge diffusion}
	\author{ Lorenzo Gavassino$^1$}
	\affiliation{
		Department of Mathematics, Vanderbilt University, Nashville, TN, USA
	}
	\author{Masoud Shokri$^2$}
	\affiliation{Institut f\"ur Theoretische Physik, 
		Johann Wolfgang Goethe--Universit\"at,
		Max-von-Laue-Str.\ 1, D--60438 Frankfurt am Main, Germany}
	\date{\today}
	
	\begin{abstract}
		We obtain stability criteria for diffusive inviscid multicomponent Israel-Stewart hydrodynamics with and without background or dynamic electromagnetic fields. Our analysis is grounded on the maximum entropy principle, and it provides stability conditions that are valid around all thermodynamic equilibria, including rotating equilibria, charged equilibria, and equilibria in a background gravitational field.
		We prove that the electromagnetic part of the information current is stable and causal by construction and, therefore, the stability criteria found for Israel-Stewart theories of hydrodynamics automatically extend to similar formulations of magnetohydrodynamics.
	\end{abstract} 
	
	\maketitle
	
	\section{Introduction}
	\label{sec:intro}
Relativistic hydrodynamics effectively describes the evolution of many-body systems in terms of a few macroscopic degrees of freedom and is used to study various physical systems in cosmology, astrophysics, condensed matter physics, and heavy-ion physics \cite{rezzolla_book,denicol_rischke_2022,andersson2007review,FlorkowskiReview2018}. 
	For simple uncharged fluids, the equations of motion are given by energy-momentum conversion but, depending on the system of interest, other single or multiple currents might exist and have to be taken into account \cite{Landry2020}. 
	In particular, a lot of interest has been given in recent years to heavy-ion collisions with lower energies, which supposedly span regions of the QCD phase diagram with finite baryon chemical potentials. 
	This growing interest has motivated numerical studies of hydrodynamics with baryon charge diffusion \cite{Denicol:2018rbw, Du:2019obx}. 
	However, in such systems, according to the underlying physics, electric charge and strangeness are also conserved, and, therefore, a multi-component hydrodynamic theory needs to be applied.
	Similar situations are hypothesized to exist in neutron stars \cite{andersson_comer2000,prix2004,sourie_glitch2017,Geo2020}.

	A variety of multi-component hydrodynamic formulations exist in the literature. 
	A non-exhaustive list includes extensions of Israel-Stewart hydrodynamics \cite{Prakash:1993bt,Monnai:2010qp,Monnai:2010th,Harutyunyan:2023nvt}, Denicol-Niemi-Molnar-Rischke (DNMR) second-order hydrodynamics \cite{Fotakis:2022usk}, and Carter's multifluid theory \cite{carter1991,Carter_starting_point,Termo}. 
	Also, the fluid might be coupled to dynamic or background electromagnetic fields, which might affect the transport of the conserved charges. 
	For example, a highly conducting fluid that starts with an electric charge separation tends to electrically neutralize. 
	In a single-component fluid, this will also lead to a vanishing baryon charge density. 
	However, in a multi-component fluid, where electric and baryon density are independent degrees of freedom, the electric neutralization due to strong electromagnetic fields might affect the baryon density, and strangeness, in nontrivial ways. 
	Understanding the evolution of such fluids requires an appropriate multi-component magnetohydrodynamic theory. 
	In the DNMR approach, a single-component resistive dissipative theory has been developed \cite{Denicol:2018rbw,Denicol:2019iyh}, which in principle can be extended similarly to Ref.\ \cite{Fotakis:2022usk}. 
	In this paper, a multi-component extension of the Israel-Stewart hydrodynamics \cite{Israel_Stewart_1979,OlsonLifsh1990} in the presence of electromagnetic fields is considered, which will be called the Israel-Stewart-Maxwell model. 
	
	In developing, and solving, dissipative relativistic hydrodynamics theories a key question is the stability of equilibrium \cite{Hiscock_Insatibility_first_order,Hishcock1983,Kovtun2019}, i.e., if the theory predicts equilibrium states to be resilient against arbitrary small perturbations.
	This question can be addressed using the so-called information current method \cite{Hishcock1983,OlsonLifsh1990,GavassinoGibbs2021}, which relies on the maximum entropy principle and does not need to assume a homogeneous equilibrium state.
	Therefore, it is valid for all thermodynamic equilibria, including those with acceleration, rotation, and background gravitational fields, and also addresses the related question of linear causality.
	A recent application of this method can be found in Ref.\ \cite{Almaalol:2022pjc} which investigates the stability of multi-component viscid Isreal-Stewart hydrodynamics. 
	In the present work, on the other hand, we apply the information-current method to derive the linear stability criteria of the multi-component \textit{inviscid} Israel-Stewart-Maxwell model, which is, in particular, relevant to numerical solutions \cite{Dash:2022xkz}.
	To this end, we start by introducing the multi-component Israel-Stewart theory in Sec.\ \ref{sec:consti-rels}.
	Then, in Sec.\ \ref{sec:gibbs}, we review Gibbs' stability criterion and derive the information current for the multi-component Israel-Stewart theory.
	The result is used in Sec.\ \ref{sec:multi-is-stabilty} to find the stability criteria.
	After considering special cases, we show that the information current of the multi-component Israel-Stewart theory can be transformed into the information current of Carter's multifluid theory.
	The stability conditions of the latter theory are known, which leads us to the multi-component Israel-Stewart theory stability conditions.
	In Sec. \ref{sec:bg-fields}, we turn to the multi-component Israel-Stewart theory with background electromagnetic fields. We show how the thermodynamic potential must be modified due to the background Lorenz force.
	Interestingly, we find that background electromagnetic fields do not alter the stability conditions of the multi-component Israel-Stewart theory.
	Finally, in Sec. \ref{sec:dynamic-fields}, we study the multi-component Israel-Stewart-Maxwell model, where electromagnetic fields are dynamic degrees of freedom, and we prove our core result: \textit{the electromagnetic part of the information-current is stable and causal by construction, and, therefore, the stability conditions of a dissipative hydrodynamic theory extend to a similar theory of resistive dissipative magnetohydrodynamics, if the medium is non-polarizable and non-magnetizable}.
    Our results, in particular, not only agree with Ref.\ \cite{Biswas:2022gwa}'s results, but also predict how they extend to resistive fluids, and to inhomogeneous equilibria.
	The paper concludes in Sec.\ \ref{sec:connclusion}.
	For the reader's convenience, details of the calculations, as well as clarifying examples, are presented in several appendices.

	\paragraph*{Notations and conventions}
	We use natural units ($\hbar=c=k_B=1$) and mostly plus metric sign convention, i.e., $\eta_{\mu \nu} = \diag(-1,1,1,1)$. 
	The covariant, exterior, and Lie derivatives are denoted by $\nabla$, $\dd$, and $\lieder{}{}$, respectively. 
	The convention for the totally antisymmetric tensor $\epsilon^{\mu\nu\a\b}$ is such that in Minkowskian coordinates $\epsilon^{0123}=-\epsilon_{0123}=1$. 
	Standard symmetrization and antisymmetrization notations are defined as $A_{(\mu\nu)} = \tfrac{1}{2}\left(A_{\mu\nu} +A_{\nu\mu}\right)$ and
	$A_{[\mu\nu]} = \tfrac{1}{2}\left(A_{\mu\nu} -A_{\nu\mu}\right)$, respectively.
	We use the notation $A^{\langle\mu\rangle} = \Delta^{\mu\nu}A_\nu$, where $\Delta^{\mu\nu}=g^{\mu\nu}+u^\mu u^\nu$, with $u^\mu$ being the fluid's four-velocity.
	$\Delta^{\mu\nu}$ projects every vector $A^\mu$ onto the plane orthogonal to $u^\mu$. 
	\section{multicomponent Israel-Stewart theory for charge diffusion}
	\label{sec:multi-is}
	
	Let us start by introducing the multi-component Israel-Stewart theory, which is an extension of the single-charge Israel-Stewart theory in the Landau frame \cite{OlsonLifsh1990}. 
	For reasons that will become clear later, it is convenient to first consider the case without electromagnetic fields. 
	We allow for an arbitrary number of conserved and non-conserved chemical species. 
	However, we neglect bulk and shear viscosity, which would make a fully general analysis unmanageable analytically.
	
	\subsection{Constitutive relations}\label{sec:consti-rels}
	
	The fluid's state is characterized by the fields $\varphi_i=\{\ed,u^\mu,n_A,V_A^\mu \}$, representing respectively the energy density, the flow velocity, the chemical densities, and the diffusive currents.
	The label $A=1,...,l$ is a chemical index. 
	Since $u^\mu$ is normalized ($u_\mu u^\mu=-1$) and $V_A^\mu$ are orthogonal to the flow lines ($u_\mu V_A^\mu=0$), the algebraic degrees of freedom are $4(1+l)$. 
	We note that in a general fluid mixture, there are no preferred chemical species. 
	Therefore the Eckart frame is not convenient and it is easier to use the Landau frame.
	The consequences of this choice are nevertheless immaterial, since in the linear regime all frames are equivalent \cite{GavassinoUniversality2023} (the only exception is the ``general frame'' \cite{NoronhaGeneralFrame2021}, which has more degrees of freedom). 
	The Landau frame entails that $u_\mu T^{\mu \nu}=-\ed u^\nu$ and $u_\mu N_A^\mu=-n_A$, so that, if we neglect bulk and shear viscosity, we have the following constitutive relations for the stress-energy tensor, particle, and the entropy current:
	\begin{subequations}
		\label{eq:const-rels}
		\begin{eqnarray}
			\label{eq:const-em-tensor}
			T^{\mu \nu}&=& (\ed+P)u^\mu u^\nu +P g^{\mu \nu} \, , \\
			\label{eq:const-current}
			N_A^\mu &=& n_A u^\mu +V_A^\mu \, , \\
			\label{eq:const-entropy}
			S^\mu &=& s u^\mu -\dfrac{\mu^A}{T} V_A^\mu - b^{AB}V_A^\nu V_{B \nu} \dfrac{u^\mu}{2T} \,,
		\end{eqnarray}
	\end{subequations}
	where Einstein summation is assumed for the chemical indices $A$ and $B$ (see Appendix \ref{app:basis} for a simple example).
	The scalars $P$, $s$, $T$, and $\mu^A$ in the constitutive relations \eqref{eq:const-rels} are the equilibrium pressure, entropy density, temperature, and chemical potentials of the fluid, respectively. 
	They are pure functions of $\ed$ and $n_A$, satisfying standard thermodynamic identities, namely the first law, the Gibbs-Duhem equation, and the Euler relation, respectively
	\begin{subequations}
		\label{eq:thermo-identities}
		\begin{align}
			\label{eq:ti-firstlaw}
			& d\ed = Tds +\mu^A dn_A \, , \\
			\label{eq:ti-gibss-dhuem}
			& dP=sdT+n_A d\mu^A \, , \\ 
			\label{eq:ti-euler}
			& \ed+P=Ts+\mu^A n_A \, .
		\end{align}    
	\end{subequations}
	The symmetric $l\times l$ matrix $b^{AB}$  in Eq.\ \eqref{eq:const-entropy} quantifies the cost in entropy density associated with diffusive currents. 
	
	\subsection{Field equations}
	\label{sec:field-eqs}

	Out of the $4(1{+}l)$ equations needed to close the system, $4{+}l$ are the balance laws for energy, momentum, and particles:
	\begin{subequations}\label{eq:eom}
		\begin{eqnarray}
			\label{eq:eom-em}
			\nabla_\mu T^{\mu \nu}&=& 0 \, , \\
			\label{eq:eom-n}
			\nabla_\mu N_A^\mu &=& R_A \, ,
		\end{eqnarray}    
	\end{subequations}
	where $R_A(\ed,n_B)$ are some reaction rates. 
	The remaining field equations are the equations of motion for $V_A^\mu$, which should be constructed to guarantee that $\nabla_\mu S^\mu \geq 0$ \cite{Israel_2009_inbook} holds \textit{exactly}. 
	In order to keep the following stability analysis as general as possible, we will not construct explicit evolution equations.  
	Any system of equations that fulfills the second law of thermodynamics is permitted.
	An example of such a system is discussed in Appendix \ref{app:eom}.
	\section{The Gibbs stability criterion}\label{sec:gibbs}
	
	Our stability analysis is grounded on the Gibbs stability criterion \cite{GavassinoGibbs2021,PrigoginebookModernThermodynamics2014,GavassinoGENERIC2022}. 
	The main idea is that, if (i) the second law of thermodynamics holds on all possible solutions of the fluid equations, i.e., $\nabla_\mu S^\mu \geq 0$ always holds, and (ii) the state of thermodynamic equilibrium maximizes the total entropy for fixed values of the conserved charges, then the total entropy is a Lyapunov function \cite{lasalle1961stability}, and the equilibrium state is Lyapunov stable. 
	If the equilibrium state maximizes the entropy in \textit{all} reference frames, then the theory is linearly causal \cite{GavassinoCausality2021}, covariantly stable, i.e, stable in all reference frames \cite{GavassinoSuperluminal2021,GavassinoBounds2023}, and symmetric hyperbolic \cite{GavassinoUniversality2023,GavassinoCasimir2023,GavassinoNonHydro2022}. 
	In this section, we briefly review this method and derive the so-called information current of the multi-component Israel-Stewart theory in the inviscid limit.
	
	\subsection{Extremum principle for open systems}\label{sec:open-extr}

	Consider an isolated system, comprised of our fluid \eqref{eq:const-rels} in weak contact with an environment, in such a way that the total entropy is the sum of the entropies of the parts, $S_{\text{tot}}=S+S_E$, and similarly for all conserved charges $Q^I_{\text{tot}}=Q^I+Q^I_E$. 
	Here, $I$ is an index that counts the conserved charges, such as the energy and baryon number. 
	These charges are fluxes of the relevant Noether currents across a chosen Cauchy surface.
	Quantities with the label ``$\, E \,$'' refer to the environment, while quantities with no label refer to the fluid. 
	Assume that the environment is always in thermodynamic equilibrium within itself, but not necessarily with the fluid, so that $S_E=S_E(Q_E^J)$ always holds.
	Then, if the environment is much larger than the fluid, i.e. $|Q^I| \ll |Q^I_{\rm tot}|$, we can expand the total entropy as follows (cf.\ Ch.\ 7 of Ref.\ \cite{huang_book})
	\begin{equation}
		S_{\text{tot}}= S+S_E(Q^I_{\text{tot}}-Q^I) \approx S+S_E(Q^I_{\text{tot}})- \dfrac{\partial S_E(Q^J_{\text{tot}})}{\partial Q^I_E} Q^I \, ,
	\end{equation}
	where Einstein summation is assumed for the charge index $I$.
	Higher order terms converge to zero in the limit $Q^I/Q^I_E \rightarrow 0$. 
	The second law of thermodynamics requires $S_{\text{tot}}$ to be non-decreasing in time.
	But since the total charges are conserved, $S_E(Q^I_{\text{tot}})$ is constant in time, so that the function $\Phi=S+\alpha^\star_I Q^I$ is also non-decreasing in time, where the numbers
	\begin{equation}\label{lecostantidiequilibrium}
		\alpha^\star_I := -  \dfrac{\partial S_E(Q^J_{\text{tot}})}{\partial Q^I_E}
	\end{equation}
	remain constant, being functions of the total conserved charges. 
	As a consequence, the fluid evolves until it reaches the state that maximizes $\Phi$ for the given values of $\alpha^\star_I$, which are fixed by the initial condition. 
	Such a state is the state of thermodynamic equilibrium identified by specific values of $\alpha^\star_I$.
	
	The function $\Phi=S+\alpha^\star_I Q^I$ can be expressed as a hydrodynamic integral. 
	In particular, if $J^{I\mu}$ are the conserved current four-vectors associated with the charges $Q^I$, and $\Sigma$ is an arbitrary Cauchy surface, then $\Phi=\int_\Sigma \phi^\mu d\Sigma_\mu$, where $d\Sigma_\mu$ is the standard-oriented surface element (see Ref.\ \cite{MTW_book}, Box 5.\ 2), and
	\begin{equation}\label{eq:info-flux}
		\phi^\mu = S^\mu +\alpha^\star_I J^{I\mu} \, .
	\end{equation}
	In our case, there are two types of conserved currents \cite{GavassinoStabilityCarter2022}. 
The first type comprises net charge currents arising from internal symmetries of the underlying theory that are conserved in all chemical reactions.
	Such currents have form $q\indices{^I ^A}N^{\mu}_A$, where $q^{IA}$ is the conserved quantum number $I$ carried by the species $A$. 
	The other currents arise from spacetime symmetries. 
	In fact, if $K^h_\nu$ are Killing vectors, then $K^h_\nu T^{\nu \mu}$ is a conserved current ($h$ is an index that counts the Killing vectors).
	The associated charges are, e.g., the energy, the linear momentum, and the angular momentum. 
	Introducing the equilibrium fugacities $\alpha_\star^A=\alpha^\star_I q^{IA}$, where $I$ runs only over the ``chemical'' charges, and the ``combined'' Killing vector $\beta^\star_\nu = \alpha^\star_h K^h_\nu$, equation \eqref{eq:info-flux} becomes
	\begin{equation}\label{eq:info-flux-killing}
		\phi^\mu =S^\mu +\alpha_\star^A N^{\mu}_A + \beta^\star_\nu T^{\nu \mu} \, .
	\end{equation}

	\subsection{Information current of the multi-component Israel-Stewart}\label{sec:grundt}
	
	Let us now turn to the derivation of the information current of the multi-component Israel-Stewart theory introduced in the previous section.
	In order to find the state that maximizes $\Phi$, we introduce a smooth one-parameter family $\varphi_i(\lambda)$ of solutions to the fluid equations, in a fixed background metric, where $\lambda=0$ is the equilibrium state. 
	Then, we write $\Phi$ as a function of $\lambda$, and we demand that $\dot{\Phi}(0)=0$ and $\ddot{\Phi}(0)\leq 0$ (with $\dot{f}=df/d\lambda$). 
	Since the equilibrium state maximizes $\Phi$ at constant $\alpha^\star_I$, the parameters $\alpha^A_\star$ and $\beta^\star_\nu$ do not depend on $\lambda$ (the ``$\star$'' helps us keep track of this distinction). 
	Then, we have
	\begin{subequations}
		\label{eq:info-c-integrand}
		\begin{eqnarray}
			\label{eq:info-c-integrand-phi}
			\phi^\mu &=& \left[s{+}\alpha_\star^A n_A{+}\beta^\star_\nu u^\nu \left(\ed{+}P\right){-}\rabbit \right]u^\mu {+} \left(\alpha_\star^A{-}\alpha^A\right)V_A^\mu {+}P\beta^{\star \mu} \, , \\
			\label{eq:info-c-integrand-phidot}
			\dot{\phi}^\mu &=&\left[s{+}\alpha_\star^A n_A{+}\beta^\star_\nu u^\nu \left(\ed{+}P\right){-}\rabbit \right]\dot{u}^\mu
			{+} \left[\dot{s}{+}\alpha_\star^A \dot{n}_A{+}\beta^\star_\nu \dot{u}^\nu \left(\ed{+}P\right) {+}\beta^\star_\nu u^\nu \left(\dot{\ed}{+}\dot{P}\right){-}\dot{\rabbit}\right]u^\mu \\
			\nonumber
			&&{+} (\alpha_\star^A{-}\alpha^A)\dot{V}_A^\mu{-}\dot{\alpha}^A V_A^\mu{+}\dot{P}\beta^{\star \mu} \, , \\
			\label{eq:info-c-integrand-phiddot}
			\ddot{\phi}^\mu &=& \left[s{+}\alpha_\star^A n_A{+}\beta^\star_\nu u^\nu (\ed{+}P){-}\rabbit \right]\ddot{u}^\mu{+} \left[\ddot{s}{+}\alpha_\star^A \ddot{n}_A{+}\beta^\star_\nu \ddot{u}^\nu (\ed{+}P) {+}\beta^\star_\nu u^\nu (\ddot{\ed}{+}\ddot{P})+2\beta^\star_\nu \dot{u}^\nu(\dot{\ed}{+}\dot{P}){-}\ddot{\rabbit}\right]u^\mu \\\nonumber
			&& {+}2 \left[\dot{s}{+}\alpha_\star^A \dot{n}_A{+}\beta^\star_\nu \dot{u}^\nu (\ed{+}P) {+}\beta^\star_\nu u^\nu (\dot{\ed}{+}\dot{P}){-}\dot{\rabbit}\right]\dot{u}^\mu  {+} (\alpha_\star^A{-}\alpha^A)\ddot{V}_A^\mu{-}\ddot{\alpha}^A V_A^\mu{-}2\dot{\alpha}^A \dot{V}_A^\mu+\ddot{P}\beta^{\star \mu} \, , 
		\end{eqnarray}
	\end{subequations}
	where we have introduced the compact notations $\alpha^A:=\mu^A/T$ and $\rabbit:=b^{AB}V_A^\nu V_{B \nu}/2T$.
	The stationary point requirement is that $\dot{\Phi}(0)=\int_\Sigma \dot{\phi}^\mu(0)d\Sigma_\mu$ must vanish for all Cauchy surfaces $\Sigma$ and for all families $\varphi_i(\lambda)$.
	This leads to the well-known equilibrium conditions,
	\begin{equation}\label{eq:equil-conditions}
		\dfrac{\mu^A}{T}=q^{IA}\alpha^\star_I \, ,  \spc \dfrac{u_\mu}{T}= \beta^\star_\mu \, , \spc V_A^\mu =0 \, .
	\end{equation}
	The first condition entails both diffusive and chemical equilibrium, the second one entails thermal and mechanical equilibrium, and the third one is simply the requirement that no irreversible flux can exist in equilibrium. 
	Plugging Eq.\ \eqref{eq:equil-conditions} into Eq.\ \eqref{eq:info-c-integrand-phiddot} and using Eq.\ \eqref{eq:thermo-identities}, we obtain
	\begin{equation}
		T\ddot{\phi}^\mu(0) =  \big[T\ddot{s}+\mu^A \ddot{n}_A+u_\nu \ddot{u}^\nu (\ed{+}P) - \ddot{\ed}-T\ddot{\rabbit}\big]u^\mu-2 \dot{P}\dot{u}^\mu  -2T\dot{\alpha}^A \dot{V}_A^\mu \, .
	\end{equation}
	In order for $\ddot{\Phi}(0)$ to be positive for all Cauchy surfaces $\Sigma$ and for all nonvanishing perturbations $\dot{\varphi}_i(0)$, the information current $E^\mu=-\ddot{\phi}^\mu(0)/2$, as is proved in the next subsection, must be a future-directed non-spacelike vector.
	In order to find $E^\mu$, we first note that
	\begin{eqnarray}
		T\ddot{\rabbit}(0)&=& b^{AB}\dot{V}_A^\nu \dot{V}_{B \nu} \, ,\qquad 
		u_\nu \ddot{u}^\nu = -\dot{u}_\nu \dot{u}^\nu \, , \qquad
		\ddot{\ed}= T \ddot{s}+\mu^A\ddot{n}_A +\dot{T} \dot{s} +\dot{\mu}^A \dot{ n}_A \, ,
	\end{eqnarray}
	which follow respectively from Eq.\ \eqref{eq:equil-conditions}, $u_\nu u^\nu=-1$, and Eq.\ \eqref{eq:thermo-identities}.
	Introducing the notation $\delta \varphi_i = \dot{\varphi}_i(0)$, and using these identities, we arrive at the formula for the information current:
	\begin{equation}\label{eq:is-info-current}
		TE^\mu = \big[ \delta T \delta s +\delta \mu^A \delta n_A +(\ed+P)\delta u^\nu \delta u_\nu + b^{AB} \delta V_A^\nu \delta V_{B\nu} \big]\dfrac{u^\mu}{2} + (s \delta T+n_A \delta \mu^A)\delta u^\mu +T\delta \bigg(\dfrac{\mu^A}{T} \bigg)\delta V_A^\mu \, .
	\end{equation}
	In appendix \ref{app:olson-ic}, we show that, for fluids with a single chemical species, i.e., $l=1$, this information current reduces to the ``energy current'' of Olson \cite{OlsonLifsh1990} in the inviscid limit. Hence, when $l=1$, the stability analysis in this work automatically reduces to Olson's analysis.
	
	\subsection{Proof that stability follows from the extremum principle}
	
	For completeness, let us review the proof that if $E^\mu$ is future-directed non-spacelike, the equilibrium state is stable \cite{GavassinoGibbs2021}.
	Let us first note that, since $\alpha^\star_I$ are constant numbers, and $J^{I\mu}$ are conserved currents, then \eqref{eq:info-flux} implies $\nabla_\mu \phi^\mu = \nabla_\mu S^\mu \geq 0$.  
	Hence, by application of the Gauss theorem, we conclude that $\Phi(\lambda)$ is non-decreasing in time, and $\Phi(0)$ is constant (being the equilibrium Massieu function \cite{Callen_book}).
	Then, using $\Phi(\lambda) = \Phi(0) + \frac{1}{2} \lambda^2\ddot{\Phi}(0)+\mathcal{O}(\lambda^3)$, and recalling that $E^\mu = -\ddot{\phi}/2$, we find that the functional
	\begin{equation}
		E= \int_\Sigma E^\mu d\Sigma_\mu= \lim_{\lambda \rightarrow 0} \dfrac{ \Phi(0)-\Phi(\lambda)}{\lambda^2} 
	\end{equation}
	is non-increasing in time.
	But if $E^\mu$ is future-directed non-spacelike, $E$ plays the role of a square integral norm of the perturbation fields $\delta \varphi_i$. 
	Hence, the equilibrium state is linearly stable \cite{Geroch_Lindblom_1991_causal}, in the sense that the linear perturbation fields $\delta \varphi_i$ evolve keeping the norm $E$ between zero and its initial value.

	\section{Stability conditions}\label{sec:multi-is-stabilty}
	
	Now that we have the information current $E^\mu$, we can derive the stability conditions for multi-component Israel-Stewart from the requirement that $E^\mu$ must be future-directed non-spacelike. 
	We will first study the fluid and diffusive sectors independently. 
	Then, using the classification of Ref.\ \cite{GavassinoUniversality2023}, we will find the change of variables that transforms the information current \eqref{eq:is-info-current} to the one of Carter's multifluid theory. 
	This will enable us to obtain the general stability conditions of the multi-component Israel-Stewart theory. 
	Finally, we will show that our results in the single component limit reproduce the existing results of \citet{OlsonRegular1990} in the inviscid limit.
	In the following, we assume $T>0$ (see \S 10 \cite{landau5} for the justification).

	\subsection{Stability of the fluid sector}\label{sec:fluid-sector}
	
	If we impose $\delta V_A^\nu=0$ into Eq.\ \eqref{eq:is-info-current}, then $E^\mu$  reduces to the information current of a perfect-fluid mixture, whose stability conditions are well known, and entail, among other things, the following constraints \cite{GavassinoGibbs2021,GavassinoCausality2021,Hishcock1983}:
	\begin{subequations}\label{eq:fluid-stability}
		\begin{eqnarray}
			\label{eq:fluid-stability-h}
			&& \ed + P > 0\,,\\
			\label{eq:fluid-stability-cs}
			&&c_s^2 = \dfrac{\partial P}{\partial \ed}\bigg|_{n_A/s} \in (0,1]\,\\
			\label{eq:fluid-stability-cv}
			&& c_v = T \dfrac{\partial s}{\partial T}\bigg|_{n_A} >0 \, . 
		\end{eqnarray}
	\end{subequations}
	The first condition guarantees stability against spontaneous acceleration, or, in other words, enforces positive inertia.
	Equation \eqref{eq:fluid-stability-cs} ensures the subluminality of adiabatic sound waves and positive adiabatic compressibility, while Eq.\ \eqref{eq:fluid-stability-cv} assures stability to heat exchanges.
	
	\subsection{Stability of the diffusive sector}\label{sec:diff-sector}
	
	We can obtain other stability conditions by focusing on perturbations that preserve both mechanical and thermal equilibrium, i.e., those with $\delta u^\nu = \delta T=0$. 
	Under these assumptions, the information current \eqref{eq:is-info-current} reduces to
	\begin{equation}\label{eq:is-info-current-diff}
		TE^\mu = \left[ \dfrac{\partial \mu^A}{\partial n_B}\bigg|_T \delta n_A \delta n_B + b^{AB} \delta V_A^\nu \delta V_{B\nu} \right]\dfrac{u^\mu}{2} +\dfrac{\partial \mu^A}{\partial n_B}\bigg|_T \delta V_A^\mu \delta n_B \,,
	\end{equation}
	where we have used $\d\mu^A=\d n_B(\partial \mu^A/{\partial n_B})_T$.
	This information current has the same structure as Carter's multifluid theory information current  (cf.\ Eq.\ (42) of \cite{GavassinoStabilityCarter2022}). 
	Hence, the resulting stability conditions are analogous: the three $l \times l$ symmetric matrices 
	\begin{equation}\label{eq:diff-stability}
		\dfrac{\partial \mu^A}{\partial n_B}\bigg|_T \, , \spc b^{AB} \, , \spc b^{AB}- \dfrac{\partial \mu^A}{\partial n_B}\bigg|_T\,, 
	\end{equation}
	are required to be positive definite for stability (actually, the third matrix is allowed to be semi-definite).
	The proof is summarised in appendix \ref{app:carter-ic}. 
	From a physical perspective, we can interpret these three stability conditions as stability to particle diffusion, stability to the spontaneous flux formation, and a causality bound on diffusion, respectively.
	Such a causality bound is the multi-component generalization of Eq.\ (77) of Ref.\ \cite{BritoDenicol2020}.

	\subsection{Systematic derivation of all stability conditions}
	
	In Secs.\ \ref{sec:fluid-sector} and \ref{sec:diff-sector}, we have provided some ``easy to check'' necessary conditions for stability. 
	Now it is time to find the complete set of necessary and sufficient conditions. 
	This seems to be a formidable task for an arbitrary $l$, considering that the information current \eqref{eq:is-info-current} is rather complicated. 
	Luckily, there is a useful trick, arising from the fact that the connection between our information current and that of Carter's theory is not accidental.
	In fact, according to the classification introduced in Ref.\ \cite{GavassinoUniversality2023}, the multi-component Israel-Stewart theory belongs to the $(l,l,0)-({\leq}l, 0,0)$ universality class, whose representative is Carter's multifluid theory \cite{carter1991,noto_rel,Carter_starting_point,langlois98,Termo,GavassinoKhalatnikov2022}. 
    Hence, there must exist a change of variables that transforms the information current \eqref{eq:is-info-current} into that of Carter's theory, whose stability conditions are known \cite{GavassinoStabilityCarter2022}. 
    Such a change of variables is
	\begin{subequations}
		\begin{eqnarray}
			\delta j_s^\mu &=&s \, \delta u^\mu -\dfrac{\mu^A}{T}\delta V_A^\mu \, , \\
			\delta j_A^\mu &=&n_A \delta u^\mu +\delta V_A^\mu \, ,
		\end{eqnarray}      
	\end{subequations}
	and its inverse is
	\begin{subequations}\label{eq:is-to-carter}
		\begin{eqnarray}
			\delta u^\mu &=& \dfrac{T\delta j_s^\mu +\mu^A \delta j_A^\mu}{\ed+P} \, , \\
			\delta V_A^\mu &=& -\dfrac{Tn_A \delta j_s^\mu}{\ed+P}+ \bigg[\delta^B_A - \dfrac{n_A \mu^B}{\ed+P} \bigg]\delta j_B^\mu \, .
		\end{eqnarray}      
	\end{subequations}
	Here, $\delta j_s^\mu$ and $\delta j_A^\mu$ are the linear perturbations to the fluxes of entropy and particles, as measured in the equilibrium local rest frame. 
 Note that $u_\mu \delta j_s^\mu=u_\mu \delta j_A^\mu =0$. 
	Introducing the ``extended'' chemical indices $X,Y \in \{s,1,...,l\}$, which treat the entropy as a ``zeroth chemical species'' (such that $n^s=s$ and $\mu^s = T$) we can write
	$
	\d \mu^X = \dermat^{XY}\d n_Y
	$,
	where $\dermat^{XY}$ a symmetric $(l+1)\times(l+1)$ matrix that can be written as a block matrix as
	\begin{equation}
		\dermat^{XY}=\dfrac{\partial \mu^X}{\partial n_Y}=
		\begin{bmatrix}
			\dfrac{\partial T}{\partial s} & & \dfrac{\partial \m}{\partial s}  \\[7pt]
			\dfrac{\partial T}{\partial \n} & & \dfrac{\partial \m}{\partial \n} \\[7pt]
		\end{bmatrix}\,.
	\end{equation}
	Here, $\n{=}[n_1,...,n_l]^T$ and $\m{=}[\mu^1,...,\mu^l]$.
	Consequently, Eq.\ \eqref{eq:is-info-current} can be rewritten in the form
	\begin{equation}\label{eq:carter-ic-form}
		TE^\mu = \big[\dermat^{XY} \delta n_X \delta n_Y + \mathcal{K}^{XY}\delta j_X^\nu \delta j_{Y\nu}  \big] \dfrac{u^\mu}{2} + \ed^{XY} \delta j_X^\mu \delta n_Y \,,
	\end{equation}
	where $\mathcal{K}^{XY}$, which is also a symmetric $(l+1)\times(l+1)$ matrix, can be written as
	\begin{equation}\label{eq:K}
		\spc \mathcal{K}^{XY}= \dfrac{1}{h^2}
		\begin{bmatrix}
			T & & \m \\
			-T \n & & h\mathbb{I}- \n \m \\
		\end{bmatrix}^T 
		\begin{bmatrix}
			h & & 0 \\
			0 & & b \\
		\end{bmatrix}
		\begin{bmatrix}
			T & & \m \\
			-T \n & & h\mathbb{I}- \n \m \\
		\end{bmatrix}\,.
	\end{equation}
	Here $h=\ed+P$, $b=[b^{AB}]$, and $\mathbb{I}$ is the $l\times l$ identity matrix.
	Note that Eq.\ \eqref{eq:K} is a congruence transformation of the matrix in the middle, arising from Eq.\ \eqref{eq:is-to-carter}.
	
	Equation \eqref{eq:carter-ic-form} is formally identical to the information current of Carter's theory, see \cite{GavassinoStabilityCarter2022} Eq.\ (42), and, therefore, it leads to the same stability conditions. 
 In particular, the fluid is stable if and only if the matrices $\ed^{XY}$, $\mathcal{K}^{XY}$, and $\mathcal{K}^{XY}-\ed^{XY}$ are positive definite. 
	Positive definiteness of $\dermat^{XY}$ guarantees thermodynamic stability in the usual ``textbook sense'' (see, e.g., \S 21 of Ref. \cite{landau5} or Ch. 12 of Ref.\ \cite{PrigoginebookModernThermodynamics2014}), positive definiteness of $\mathcal{K}^{XY}$ guarantees that all chemical components have positive kinetic energy, and positive (semi-)definiteness of $\mathcal{K}^{XY}-\dermat^{XY}$ enforces causality.
 
	\subsection{A concrete example}
	
	Let us apply the results of the previous subsection to the case $l=1$, which has already been analyzed elsewhere \cite{Hishcock1983,OlsonLifsh1990,OlsonRegular1990}. 
 In this way, we can corroborate our findings by comparing them with established literature.
	The matrix $\dermat^{XY}$ is $2{\times}2$ now, and its first diagonal entry is $T/c_v$, which is automatically positive if condition \eqref{eq:fluid-stability} holds.
	Hence, we only need to compute the determinant of $\dermat^{XY}$:
	\begin{equation}
		\det[\ed^{XY}]= \dfrac{\partial(T,\mu)}{\partial(s,n)} = \dfrac{\partial(T,\mu)}{\partial(T,n)} \dfrac{\partial(T,n)}{\partial(s,n)} = \dfrac{\partial \mu}{\partial n}\bigg|_T \dfrac{\partial T}{\partial s}\bigg|_n \,,
	\end{equation}
	which is positive if both conditions \eqref{eq:fluid-stability} and \eqref{eq:diff-stability} are fulfilled.
	Consequently, conditions \eqref{eq:fluid-stability} and \eqref{eq:diff-stability} guarantee that $\dermat^{XY}$ is positive definite. 
	Furthermore, one can verify that $\mathcal{K}^{XY}$ is positive definite if and only if $\ed+P$ and $b$ are positive, which are also ensured if conditions \eqref{eq:fluid-stability} and \eqref{eq:diff-stability} are satisfied.
	Thus, we are left only with the problem of determining the causality constraints arising from the positive definiteness of $\mathcal{K}^{XY}{-}\ed^{XY}$. 
	Rather than studying this matrix directly, it is more convenient to first perform the following congruence transformation, which diagonalizes $\mcK$ preserving positive definiteness:
	\begin{equation}\label{eq:awry}
		\begin{bmatrix}
			s & -\mu/T \\
			n &  1 \\
		\end{bmatrix}^T
		[\mathcal{K}^{XY}{-}\ed^{XY}]
		\begin{bmatrix}
			s & -\mu/T \\
			n &  1 \\
		\end{bmatrix}
		=
		\begin{bmatrix}
			h(1{-}c_s^2) & & -nT \dfrac{\partial \alpha}{\partial n}\bigg|_{\mathfrak{s}} \\[7pt]
			-nT \dfrac{\partial \alpha}{\partial n}\bigg|_{\mathfrak{s}} & & b-T \dfrac{\partial \alpha}{\partial n}\bigg|_{\ed}\\[7pt]
		\end{bmatrix}\,,
	\end{equation}
	where we have used in particular
	\begin{equation}
		\pdv{X}{n}\bigg|_s = \pdv{X}{n}\bigg|_\mathfrak{s} - \mathfrak{s}\pdv{X}{s}\bigg|_n\,,
		\qquad
		\pdv{P}{n}\bigg|_\mathfrak{s} = \frac{h}{n}\pdv{P}{\ed}\bigg|_\mathfrak{s}\,,
		\qquad
		T\pdv{\alpha}{n}\bigg|_s - \mu\pdv{\alpha}{s}\bigg|_n = T\pdv{\alpha}{n}\bigg|_\ed\,,
	\end{equation}
	with $X\in\{\alpha, P\}$, and where we recall that $h=\ed+P$, $\alpha=\mu/T$, and we have defined $\mathfrak{s}=s/n$ (the specific entropy).
	The positive definiteness of the first diagonal entry implies $c_s^2 <1$, which agrees with condition \eqref{eq:fluid-stability}. 
	The positivity of the second diagonal entry leads to a more stringent version of the positivity condition on the third matrix in \eqref{eq:diff-stability}. 
	In particular, we have that
	\begin{equation}\label{eq:single-is-sc}
		b> T\dfrac{\partial \alpha}{\partial n}\bigg|_\ed = \dfrac{\partial \mu}{\partial n}\bigg|_T +T^3 \dfrac{\partial T}{\partial \ed}\bigg|_{n} \bigg( \dfrac{\partial \alpha}{\partial T}\bigg|_n \bigg)^2 \geq \dfrac{\partial \mu}{\partial n}\bigg|_T \, , 
	\end{equation}
	where we have used Eqs.\ (82) and (97) of Ref.\ \cite{Hishcock1983}, namely
	\begin{equation}
		\pdv{T}{n}\bigg|_\ed = T^2\pdv{\alpha}{\ed}\bigg|_n\,,\qquad
		\pdv{T}{\ed}\bigg|_n  \geq 0\,.
	\end{equation}
	However, the most stringent causality condition comes from the positivity of the determinant of the matrix \eqref{eq:awry}. 
	With the aid of the identities
	\begin{subequations}
		\begin{eqnarray}
			nT\dfrac{\partial \alpha}{\partial n}\bigg|_{\mathfrak{s}} &=& \dfrac{hc_s^2}{n} - \dfrac{h}{T}\dfrac{\partial T}{\partial n}\bigg|_{\mathfrak{s}} \, , \\
			nT\dfrac{\partial \alpha}{\partial n}\bigg|_{\ed} &=& \dfrac{hc_s^2}{n} - \dfrac{2h}{T}\dfrac{\partial T}{\partial n}\bigg|_{\mathfrak{s}} + \dfrac{h^2}{n^2 T^2} \dfrac{\partial T}{\partial \mathfrak{s}}\bigg|_n \, ,
		\end{eqnarray}
	\end{subequations}
	we obtain the following causality constraint:
	\begin{equation}\label{eq:single-is-cc}
		h(1-c_s^2)\bigg[ \dfrac{n^2 b +h}{h^2}-\dfrac{1}{nT^2} \dfrac{\partial T}{\partial \mathfrak{s}}\bigg|_n  \bigg]-\bigg[1-\dfrac{n}{T}\dfrac{\partial T}{\partial n}\bigg|_{\mathfrak{s}} \bigg]^2 >0 \,,
	\end{equation}
	which is precisely the causality condition of the inviscid Israel-Stewart reported by \citet{OlsonRegular1990} (see $\Omega_3$), expressed in the Landau frame through Eq.\ (88) of Ref.\ \cite{GavassinoNonHydro2022}. 
  This completes the stability analysis for the $l=1$ case.
	
	\section{Background electromagnetic fields}
	\label{sec:bg-fields}
	
	It is finally time to study the effect of electromagnetic fields on fluid stability. 
	We call the ``Israel-Stewart-Maxwell model'' a fluid whose constitutive relations are exactly the same as introduced in Sec.\ \ref{sec:consti-rels}, but the energy-momentum conservation \eqref{eq:eom-em} is replaced by the Lorentz force,
	\begin{equation}\label{eq:lorentz}
		\nabla_\mu T^{\mu \nu} =  F\indices{^\nu _\mu }J^{e\mu}= q^{eA} F\indices{^\nu _\mu }N_A^{\mu} \, ,
	\end{equation}
	where $F_{\mu\nu}$ is the Faraday tensor, and $q^{eA}$ is the electric charge of the particle of type $A$ (the electric current $J^{e\mu}=q^{eA}N_A^\mu$ is one of the conserved currents $J^{I\mu}$ introduced in Sec.\ \ref{sec:open-extr}). 
	A fluid of this kind is minimally coupled to electromagnetic fields.
	Namely, it is neither polarizable nor magnetizable, and, therefore, there are no electromagnetic corrections to the constitutive relations \eqref{eq:const-rels}, and the Israel-Stewart part communicates with the Maxwell part solely through \eqref{eq:lorentz} and the equations of motion of $V_A^\mu$. 
	More general cases are left for future investigation. 
	
	In this section, we focus on the case where electromagnetic fields are externally generated and can be treated as fixed background fields. 
	This approach is only applicable to weakly charged fluids ($J^{e\mu}\rightarrow 0$) close to strong electromagnetic sources ($F^{\alpha \beta}_{\text{ext}}\rightarrow \infty$) so that the Lorentz force is a ``$\, 0 \times \infty \,$'' indeterminate form with a finite value, but the fluid does not generate electromagnetic fields of its own. 
	
	\subsection{Conditions for the existence of equilibrium}
	
	Not all background fields $F^{\alpha \beta}$ allow an equilibrium state to form.
	An external source that fluctuates will prevent the fluid from relaxing to a stationary state. 
	Hence, it is reasonable to demand that $F^{\alpha \beta}$ be invariant under the symmetry group generated by the thermal Killing vector $\beta^{\star \mu}$ in equilibrium, i.e., $\lieder{\beta^\star}{\vb{F}}=0$, where $\vb{F}$ is the Faraday two-form. 
	Using Cartan's magic formula
	$\lieder{\beta^\star}{\vb{F}} = \beta^\star\cdot\dd{\vb{F}}+\dd \left(\beta^\star\cdot\vb{F}\right)$ (see, e.g., \cite{schutz1980}), where $\dd$ is the exterior derivative, and recalling that $\dd\vb{F}=0$, we find $\dd \left(\beta^\star\cdot\vb{F}\right)=0$.
	Consequently,
	\begin{equation}\label{eq:betaffuzzo}
		\beta^\star_\nu F\indices{^\nu _\mu}  = -\nabla_\mu \psi^\star \, ,
	\end{equation} 
	for some externally fixed background scalar potential $\psi^\star$.
	Note that, since $F^{\alpha \beta}$ is skew-symmetric, we have $\beta^{\star \mu}\nabla_\mu \psi^\star=0$, meaning that $\psi^\star$ is itself stationary in the equilibrium rest frame.
	Under assumption \eqref{eq:betaffuzzo}, conservation of energy is restored, but one needs to add the ``electrostatic potential contribution''. 
	In particular, if we contract Eq.\ \eqref{eq:lorentz} with $\beta^\star_\nu$, and use Eq.\ \eqref{eq:betaffuzzo}, together with the conservation of the electric current ($\nabla_\mu J^{e\mu}=0$), and the Killing condition $\nabla_{(\mu} \beta^\star_{\nu)}=0$, we obtain the following conservation law:
	\begin{equation}\label{energycurrentUe}
		\nabla_\mu (\beta^\star_\nu T^{\nu \mu} + \psi^\star J^{e\mu})=0 \, .
	\end{equation}
	The vector field in the brackets can be seen as the energy current of the system, which accounts for the familiar term ``electric potential $\times$ charge'' in electrostatics. 
	Note that, in Eq.\ \eqref{eq:betaffuzzo}, the vector $\beta^\star_\nu F^{\nu \mu}$ is proportional to the electric field as measured in the equilibrium local rest frame of the fluid. 
	Hence, if we set $\psi^\star=0$, we obtain a pure magnetic field and no electric field in the rest frame. 
	In this case, the conserved energy current in \eqref{energycurrentUe} reduces to the same energy current we have for vanishing electromagnetic fields. 
	
	\subsection{Equilibrium states and their stability}
	
	If we retrace, in the presence of background electromagnetic fields, analogous steps as in Sec.\ \ref{sec:gibbs}, still assuming that $\nabla_\mu S^\mu \geq 0$, we see that the result is almost the same. 
	The only difference appears in the steps between equations \eqref{eq:info-flux} and \eqref{eq:info-flux-killing}: we need to replace the current $\beta^\star_\nu T^{\nu \mu}$, which is no longer conserved, with $\beta^\star_\nu T^{\nu \mu} + \psi^\star J^{e\mu}$. 
	Hence, the current $\phi^\mu$, as defined in \eqref{eq:info-flux}, now explicitly reads
	\begin{equation}
		\phi^\mu =S^\mu +(\alpha^\star_I q^{IA}+\psi^\star q^{eA}) N^{\mu}_A + \beta^\star_\nu T^{\nu \mu} \, .
	\end{equation}
	Interestingly, this current can still be rewritten in the form \eqref{eq:info-flux-killing}, with the difference that now $\alpha^A_\star=\alpha^\star_I q^{IA}+\psi^\star q^{eA}$. 
	But then, if we repeat the procedure of Sec.\ \ref{sec:grundt}, keeping $\psi^\star$ independent of $\lambda$ ($\psi^\star$ being a background quantity), we obtain the same information current, and therefore the same stability conditions.
	In particular, the analysis of Sec.\ \ref{sec:multi-is-stabilty} is not altered. 
	The only difference is that now, in Eq. \eqref{eq:equil-conditions}, the condition of equilibrium against diffusion needs to be modified as
	\begin{equation}\label{eq:fug-bg-contr}
		\dfrac{\mu^A}{T} =\alpha^\star_I q^{IA}+\psi^\star q^{eA} \, .
	\end{equation}
	Now, if we focus on individual spacetime events, nothing has changed since we can ``reabsorb'' the electromagnetic correction into the fugacity of the electric charge (namely $\alpha^\star_e+\psi^\star \rightarrow \bar{\alpha}^\star_e$). 
	This implies that chemical equilibrium still holds. 
	There is, however, a crucial difference compared to Eq.\ \eqref{eq:equil-conditions}: while $\alpha^\star_I$ are constants (cf.\ Eq.\ \eqref{lecostantidiequilibrium}), $\psi^\star$ can instead exhibit gradients. 
	Indeed, Eq.\ \eqref{eq:fug-bg-contr} tells us that the fluid tends to \textit{stratify} to counterbalance the electric force:
	\begin{equation}
		\nabla_\mu (\mu^A/T) + q^{eA}\beta^\star_\nu F\indices{^\nu _\mu}=0 \, .
	\end{equation}
	Nevertheless, the stability conditions for these equilibrium states are unchanged. 
	In particular, the $(1{+}l)\times(1{+}l)$ symmetric matrices $\ed^{XY}$, $\mathcal{K}^{XY}$, and $\mathcal{K}^{XY}{-}\ed^{XY}$ must be positive definite.
	
	\section{Dynamical electromagnetic fields}
	\label{sec:dynamic-fields}
	
	We can finally study the ``complete'' Israel-Stewart-Maxwell model, where the Faraday tensor $F^{\mu \nu}$ is a dynamic degree of freedom. 
	Now, Eq.\ \eqref{eq:lorentz} still holds, but the system regains a conserved stress-energy tensor $T^{\mu \nu}_{\text{tot}}=T^{\mu \nu}+T^{\mu\nu}_{\rm em}$, where $T^{\mu \nu}$ is the fluid stress-energy tensor given in Eq.\ \eqref{eq:const-em-tensor}, and
	\begin{eqnarray}
		T^{\mu\nu}_{\rm em} = F^{\mu}_{\,\alpha}F^{\nu\alpha} - \frac{1}{4}g^{\mu\nu}F^{\alpha \beta}F_{\alpha \beta} \, ,
	\end{eqnarray}
	is Maxwell's stress-energy tensor, which keeps the vacuum form since the medium is not polarizable or magnetizable. 
	Note that, since the inhomogenous Maxwell equations read $\nabla_\nu F^{\mu \nu}=J^{e\mu}$ \cite{carroll_2019}, we have $\nabla_\mu T^{\mu\nu}_{\rm em} = -F\indices{^\nu _\mu }J^{e\mu}$ and, using Eq.\ \eqref{eq:lorentz}, we indeed recover the conservation law $\nabla_\mu T^{\mu\nu}_{\rm tot} = 0$.
	The constitutive relations for the entropy current and the particle currents still are given by Eq.\ \eqref{eq:const-rels}, unaffected by electromagnetic fields.
	
	\subsection{Some preliminaries}
	
	The analysis of Sec.\ \ref{sec:open-extr} also holds for the conglomerate Israel-Stewart-Maxwell fluid. 
	In this case, the thermodynamic potential $\Phi=S+\alpha^\star_I Q^I$ splits as the sum of a matter part and an electromagnetic part, $\Phi=\Phi_{\text{mat}}+\Phi_{\text{em}}$, where the first is the flux of the current \eqref{eq:info-flux-killing} (with $\alpha_\star^A=\alpha^\star_I q^{IA}$), while the second is the flux of the current
	\begin{eqnarray}
		\phi^\mu_{\text{em}}=  \beta^\star_\nu T^{\nu \mu}_{\text{em}}\, .
	\end{eqnarray}
	It is well known that the Maxwell stress-energy tensor obeys the dominant energy condition \cite{Nungesser2009}, so that $\beta^\star_\nu T^{\nu \mu}_{\text{em}}$ is a past-directed non-spacelike vector. 
	Consequently, $\Phi_{\text{em}} = \int_\Sigma \phi^\mu_{\text{em}}d{\Sigma}_\mu\leq 0$, being zero only for $F^{\mu \nu}=0$.
	
	Given that $\Phi_{\text{mat}}$ depends only on matter fields, and $\Phi_{\text{em}}$ depends only on electromagnetic fields, intuition may suggest that the maximum of $\Phi$ can be obtained by simply maximizing $\Phi_{\text{mat}}$ and $\Phi_{\text{em}}$ separately. 
 However, this would be a mistake. 
	The problem is that we need to maximize $\Phi_{\text{mat}}+\Phi_{\text{em}}$ within the space of all the \textit{physically permissible} field configurations on $\Sigma$, i.e., only those field configurations that satisfy the Maxwell equations,
	\begin{equation}\label{eq:maxwell}
		\epsilon^{\mu\nu\a\b}\nabla_\nu F_{\a\b} = 0\,,
		\qquad
		\nabla_\nu F^{\mu\nu} = J^{e\mu}\,.
	\end{equation}
	For example, calling $\mathrm{n}_\nu$ the unit normal to the hypersurface $\Sigma$, we have the following constraint:  
	\begin{equation}
		\mathrm{n}_\nu (\nabla_\mu F^{\mu \nu}+J^{e\nu})=0 \, .
	\end{equation}
	We note that, due to the anti-symmetry of $F^{\mu \nu}$, this is not an equation of motion. 
	It is a differential equation that relates $F^{\mu \nu}$ and $J^{e\nu}$ across the Cauchy surface $\Sigma$ (it is the equation ``divergence of electric field'' $=$ ``charge density''). 
	Hence, the potential $\Phi_{\text{mat}}+\Phi_{\text{em}}$ needs to be maximized within a constrained manifold of states on $\Sigma$. 
	This makes the stability analysis more delicate, but it can still be carried out rigorously, as we show below. 
	
	\subsection{Electromagnetic contributions to the information current}
	
	Following the same steps as in Sec.\ \ref{sec:grundt}, let us differentiate $\phi^\mu_{\text{em}}$ twice with respect to $\lambda$:
	\begin{subequations}
		\begin{eqnarray}
			\label{eq:em-phi-step-0}
			\phi^\mu_{\text{em}} &=& \beta^\star_\nu F^{\mu}_{\,\alpha}F^{\nu\alpha} - \frac{1}{4}F^{\alpha \beta}F_{\alpha \beta} \beta^{\star \mu}  \, ,\\
			\label{eq:em-phidot-step-0}
			\dot{\phi}^\mu_{\text{em}} &=& \beta^\star_\nu \dot{F}^{\mu}_{\,\alpha}F^{\nu\alpha}+\beta^\star_\nu F^{\mu}_{\,\alpha}\dot{F}^{\nu\alpha} - \frac{1}{2}F^{\alpha \beta}\dot{F}_{\alpha \beta} \beta^{\star \mu}  \, ,\\
			\label{eq:em-phiddot-step-0}
			\ddot{\phi}^\mu_{\text{em}} &=&  \beta^\star_\nu \ddot{F}^{\mu}_{\,\alpha}F^{\nu\alpha}+\beta^\star_\nu F^{\mu}_{\,\alpha}\ddot{F}^{\nu\alpha} - \frac{1}{2}F^{\alpha \beta}\ddot{F}_{\alpha \beta} \beta^{\star \mu} +2\beta^\star_\nu \dot{F}^{\mu}_{\,\alpha}\dot{F}^{\nu\alpha}-\frac{1}{2}\dot{F}^{\alpha \beta}\dot{F}_{\alpha \beta} \beta^{\star \mu} \, .
		\end{eqnarray}
	\end{subequations}
	Using the Maxwell equations \eqref{eq:maxwell} and $F_{\mu \nu}{=}\nabla_\mu A_\nu {-}\nabla_\nu A_\mu$, where $A^\mu$ is the electromagnetic four-potential, as is shown in Appendix \ref{app:mirable}, we rewrite $\dot{\phi}^\mu_{\text{em}}$ and $\ddot{\phi}^\mu_{\text{em}}$ as follows:
	\begin{subequations}
		\label{eq:mirable}
		\begin{eqnarray}
			\label{eq:mirable-dot}
			\dot{\phi}^\mu_{\text{em}} &=& \beta^\star_\nu A^\nu \dot{J}^{e\mu} + 2\dot{A}_\nu \beta^{\star [\nu}J^{e\mu]}+\dot{F}^{\mu \alpha} (\mathcal{L}_{\beta^\star}A)_\alpha - \dot{A}_\alpha (\mathcal{L}_{\beta^\star}F)^{\mu \alpha}+\nabla_\alpha Z_{(1)}^{[\alpha \mu]}  \, ,\\   
			\label{eq:mirable-ddot}
			\ddot{\phi}^\mu_{\text{em}} &=& \beta^\star_\nu A^\nu \ddot{J}^{e\mu} + 2\ddot{A}_\nu \beta^{\star [\nu}J^{e\mu]}+\ddot{F}^{\mu \alpha} (\mathcal{L}_{\beta^\star}A)_\alpha - \ddot{A}_\alpha (\mathcal{L}_{\beta^\star}F)^{\mu \alpha}+\nabla_\alpha Z_{(2)}^{[\alpha \mu]}  +2\beta^\star_\nu \dot{F}^{\mu}_{\,\alpha}\dot{F}^{\nu\alpha}-\frac{1}{2}\dot{F}^{\alpha \beta}\dot{F}_{\alpha \beta} \beta^{\star \mu} \, .
		\end{eqnarray}
	\end{subequations}
	Here, $\nabla_\alpha Z_{(i)}^{[\alpha \mu]}$ are some residual terms (cf.\ Eqs.\ \eqref{eq:z1} and \eqref{eq:z2}) which, when integrated over the Cauchy surface, can be transformed (by Gauss' theorem \cite{Poisson_notes}) into boundary integrals, $\int_\Sigma \nabla_\alpha Z_{(i)}^{[\alpha \mu]} d\Sigma_\mu =\frac{1}{2} \oint_{\partial \Sigma} Z_{(i)}^{[\alpha \mu]}dS_{\mu \alpha}$, which vanish if the fields decay fast enough at infinity. Hence, we can ignore these terms. 
	Consequently, for $\dot{\Phi}$ to vanish, one must have $\mathcal{L}_{\beta^\star}F=0$ in equilibrium.
 If we choose a gauge such that $\mathcal{L}_{\beta^\star}A=0$ (in equilibrium), we can then combine Eq.\ \eqref{eq:mirable} with Eq.\ \eqref{eq:info-c-integrand}, to obtain the following equilibrium conditions:
	\begin{equation}\label{eq:pastaalsugo33}
		\dfrac{\mu^A}{T}=q^{IA}\alpha^\star_I+q^{eA} \beta^\star_\nu A^\nu \, ,  \spc \dfrac{u_\mu}{T}= \beta^\star_\mu \, , \spc V_A^\mu =0 \, ,
	\end{equation}
	which are in perfect agreement with statistical mechanics \cite{Jensen2013}. 
	In a different gauge, $\tilde{A}_\alpha = A_\alpha +\nabla_\alpha \chi$, we have $(\mathcal{L}_{\beta^\star}\tilde{A})_\alpha=\nabla_\alpha (\beta^{\star \nu}\nabla_\nu \chi)$. 
	As a result, the two terms $\beta^\star_\nu \tilde{A}^\nu \dot{J}^{e\mu}$ and $\dot{F}^{\mu \alpha} (\mathcal{L}_{\beta^\star}\tilde{A})_\alpha$ in Eq.\ \eqref{eq:mirable-dot} combine together to give $\beta^\star_\nu (\tilde{A}^\nu {-}\nabla^\nu \chi) \dot{J}^{e\mu}+\nabla_\alpha (\dot{F}^{\mu \alpha} \beta^{\star \nu}\nabla_\nu \chi)$. 
	The pure divergence can be reabsorbed into $\nabla_\alpha Z_{(1)}^{\alpha \mu}$, and we can ignore it. 
	Then, the first equation of \eqref{eq:pastaalsugo33} becomes the gauge-invariant expression
	\begin{equation}
		\dfrac{\mu^A}{T}=q^{IA}\alpha^\star_I+q^{eA}  (\beta^\star_\nu \tilde{A}^\nu + \Lambda) \, ,
	\end{equation}
	where we have defined $\Lambda=-\beta^\star_\nu \nabla^\nu \chi$. 
	This formula generalizes Eq.\ (2.3) of Ref.\ \cite{HernandezKovtun2017} to multi-component systems.
	
	Finally, the information current naturally splits as the sum of Eq.\ \eqref{eq:is-info-current} and
	\begin{eqnarray}\label{eq:carbonara-with-onions}
		TE_{\text{em}}^\mu = -u_\nu \bigg[  \delta F^{\mu}_{\,\alpha}  \delta F^{\nu\alpha}-\frac{1}{4} g^{\mu \nu} \delta F^{\alpha \beta}  \delta F_{\alpha \beta}  \bigg] \, .
	\end{eqnarray}
	But the term in the brackets is just the Maxwell stress-energy tensor with $F^{\mu \nu}$ replaced by $\delta F^{\mu \nu}$. 
Hence, the dominant energy condition still applies, meaning that $E_{\text{em}}^\mu$ is future-directed non-spacelike by construction. 
In conclusion, the same stability analysis we carried out in Sec.\ \ref{sec:multi-is-stabilty} also applies to the full Israel-Stewart-Maxwell model. 
	In other words, all equilibrium states (also charged equilibria) are stable provided that the matrices $\ed^{XY}$, $\mathcal{K}^{XY}$, and $\mathcal{K}^{XY}{-}\ed^{XY}$ are positive definite. 
	This completes our analysis.

	\section{Conclusions}
	\label{sec:connclusion}
	In this work, we investigated the stability of diffusive inviscid multi-component Israel-Stewart hydrodynamics using the Gibbs stability criterion. 
	To this end, applying results of Ref.\ \cite{GavassinoUniversality2023}, we utilized the formal equivalence between the information current of multi-component Israel-Stewart theory and that of Carter's multifluid theory. 
    By means of an appropriate variable transformation, we could map the already known stability criteria of Carter's theory \cite{GavassinoStabilityCarter2022} into the stability criteria of multi-component Israel-Stewart hydrodynamics. 
	We also showed that our results reproduce the results of \citet{OlsonLifsh1990} in the single component case.
	
	Then, we extended our analysis to what we call the ``multi-component Israel-Stewart-Maxwell theory'', i.e., the Israel-Stewart theory in the presence of electromagnetic fields. 
	We first considered background electromagnetic fields and found that the thermodynamic potential is modified to account for an electrostatic potential correction. 
	Such a modification can be reabsorbed into the definition of the fugacities, which then develop gradients to counteract the electric force. 
    Interestingly, this does not affect the final formula of the information current, so the stability criteria of multi-component Israel-Stewart still hold (unchanged) in the presence of background electromagnetic fields. 
	
	Finally, we analyzed the Israel-Stewart-Maxwell theory with dynamic electromagnetic fields. 
	We showed that, for an equilibrium state to exist, the gauge field's comoving temporal component contributes to the fugacities. 
	This is a multi-component extension of well-known statistical mechanics results, which led us to our main finding: \textit{the electromagnetic part of the information current is stable and causal by construction}. 
	Consequently, the stability criteria of diffusive inviscid Israel-Stewart hydrodynamics automatically extend to the Israel-Stewart-Maxwell model, namely to diffusive resistive inviscid Israel-Stewart magnetohydrodynamics.
	
	The stability criteria derived in this work are valid for an arbitrary number of chemical species, with or without chemical reactions. 
    Under such criteria, we could guarantee the stability of all thermodynamic equilibrium states, including rotating equilibria, globally charged equilibria, and equilibria in the presence of background gravitational fields. 
    Furthermore, we could automatically enforce causality \cite{GavassinoCausality2021,GavassinoSuperluminal2021,GavassinoBounds2023}, and even symmetric hyperbolicity (if the equilibrium state is homogeneous \cite{GavassinoUniversality2023}).
	
	In this work, we have neglected the effects of bulk and shear viscosities. 
	However, our main finding that the electromagnetic sector is stable will hold also in that case. 
    Hence, the stability criteria of diffusive viscid Israel-Stewart hydrodynamics automatically extend to the corresponding formulation of magnetohydrodynamics. 
	Finding the stability criteria of such a theory in the multi-component case is a complicated task that, if doable, is a natural extension of the present work. 
	We have also assumed that the medium is non-polarizable and non-magnetizable. 
	Assessing the stability of polarizable and magnetizable media will be a worthwhile extension of the current work.

	\section*{Acknowledgements}
	
	This research was supported in part by the National Science Foundation under Grant No. PHY-1748958.
	L.G. is partially supported by a Vanderbilt's Seeding Success Grant. 
	M.S. is supported by the Deutsche Forschungsgemeinschaft
	(DFG, German Research Foundation) through the Collaborative Research
	Center CRC-TR 211 ``Strong-interaction matter under extreme conditions''
	- project number 315477589 - TRR 211 and by the State of Hesse within the Research Cluster
	ELEMENTS (Project ID 500/10.006).
	
	\appendix
	
	\section{An example of basis transformation}
	\label{app:basis}
	
	Taking inspiration from the physics of heavy-ion collisions (see Ref.\ \cite{Fotakis:2022usk} and references therein), let us consider (as a simple example) a fluid comprised of three flavours of quarks. 
	Then, we can work in the so-called flavour basis, for which $A=\{u,d,s\}$, referring to respectively  ``up'', ``down'', and ``strange'' quarks. 
	Consequently, the electric charge carried by each chemical species is $q^{eA}=[2/3,-1/3,-1/3]^T$.
	This system can be equivalently described in the so-called $BQS$ (namely ``baryon'', ``electric'', and ``strangeness'') basis, defined as follows:  $n_B = (n_u+n_d+n_s)/3$, $n_Q = 2n_u/3-(n_d+n_s)/3$, and $n_S=-n_s$. These relations can be compactly expressed using chemical index notation \cite{Carter_starting_point} as
	\begin{equation}
		\tilde{n}_A = L\indices{^B_A} n_B  \,,
	\end{equation}
	where
	\begin{eqnarray}
		L = \begin{bmatrix}
			
			1/3 & 2/3 & 0 \\
			1/3 & -1/3 & 0 \\
			1/3 & -1/3 & -1
		\end{bmatrix}\,.
	\end{eqnarray}
	The chemical potentials $\mu^A$ transform with the inverse of $L$, namely
	\begin{equation}\label{eq:tildemu}
		\tilde{\mu}^A =   \left(L^{-1}\right)\indices{^A _B} \mu^B\,,
	\end{equation}
	where 
	\begin{eqnarray}
		L^{-1} = \begin{bmatrix}
			1 & 2 & 0 \\
			1 & -1 & 0 \\
			0 & 1 & -1
		\end{bmatrix}\,.
	\end{eqnarray}
	To prove this, we start from Eq.\ \eqref{eq:ti-firstlaw}, expressed in the new chemical basis, and perform the following manipulations to get to the old basis:
	\begin{equation}
		d\ed = Tds +\tilde{\mu}^A d\tilde{n}_A= Tds +\tilde{\mu}^A d(L\indices{^B_A} n_B)=Tds +(\tilde{\mu}^A L\indices{^B_A})d n_B=Tds +\mu^B dn_B \,,
	\end{equation}
	which implies $\mu^B=\tilde{\mu}^A L\indices{^B_A}$, or, equivalently, Eq.\ \eqref{eq:tildemu}.
	Note that, for this procedure to work, the matrix elements $L\indices{^B_A}$ need to be constants.
	
	Other quantities transform with $L^{-1}$ for each upper index and with $L$ for each lower index, just like in usual linear algebra.
	For example, $\tilde{q}^e = L^{-1}\cdot q^e = [0,1,0]^T$, as expected. 
	Also, this transformation law can be proved explicitly:
	\begin{equation}
		J^{e\mu}=\tilde{q}^{eA} \tilde{N}_A^\mu=\tilde{q}^{eA} L\indices{^B_A} N_B^\mu=q^{eA} N_A^\mu \quad \Longrightarrow \quad \tilde{q}^{eA} L\indices{^B_A} = q^{eB} \, .
	\end{equation}
	
	\section{Simple equations of motion}\label{app:eom}
	
	In this appendix, we work out a simple equation of motion for the diffusion currents in the presence of electromagnetic fields from the second law of thermodynamics. For simplicity, here we assume that there are no chemical reactions, namely $R_A=0$ in equation \eqref{eq:eom-n} so that all particle currents are conserved.
	Then, taking the divergence of entropy current \eqref{eq:const-entropy}, and using Eq.\ \eqref{eq:lorentz}, we find 
	\begin{eqnarray}\label{TdS}
		T\nabla_\mu S^\mu = V_{A}^{\mu}
		\left[
		q^{eA} \mcE_\mu - T\nabla_\mu \alpha^A - b^{AB} u^\nu\nabla_\nu V_{B\mu} - V_{B\mu} T \nabla_\nu\left(\frac{b^{AB}u^\nu}{2T}\right)
		\right]  \;,
	\end{eqnarray}
	where $\mcE^\mu = F^{\mu\nu}u_\nu$ is the electric four-vector.
	Now we assume 
	\begin{equation}\label{TdS2}
		T \nabla_\mu S^\mu = T V_{A}^\mu (\kappa^{-1})^{AB} V_{B\mu}  \, , 
	\end{equation}
	where $(\kappa^{-1})^{AB}$ is a positive definite, and, subsequently, the second law is guaranteed.
	We are then naturally led to postulate
	\begin{equation}\label{eq:gringo}
		q^{eA} \mcE_\mu - T\nabla_{\langle \mu \rangle} \alpha^A - b^{AB} u^\nu \nabla_\nu V_{B\langle \mu \rangle} -  T  V_{B\mu}\nabla_\nu\left(\frac{b^{AB}u^\nu}{2T}\right) = T(\kappa^{-1})^{AB} V_{B\mu} \, , 
	\end{equation}
	where $(\kappa^{-1})^{AB}$ is now recognized as the inverse of the diffusion matrix $\kappa_{AB}$, which is symmetric, by the Onsager principle, in addition to being positive definite. 
	Neglecting the last term on the left-hand side of Eq.\ \eqref{eq:gringo}, which is of higher order\footnote{Neglecting this term formally breaks the second law of thermodynamics as an exact inequality, which  no longer holds for arbitrary solutions. However, this term would anyway vanish in the linear regime, meaning that it cannot affect linear stability.} (and unphysical \cite{GavassinoFarFromBulk2023}), we obtain a system of coupled Cattaneo-type equations,
	\begin{equation}\label{eq:cattaneo}
		\tau^B_A u^\nu \nabla_\nu V_{B\langle \mu \rangle} + V_{A\mu} = -\kappa_{AB}\nabla_{\langle\mu \rangle} \alpha^B + \sigma_A \mcE_\mu\,.
	\end{equation}
	Here
	\begin{equation}
		\tau_A^B = \dfrac{\kappa_{AC} b^{CB}}{T} \, , \quad \quad \sigma_A = \dfrac{\kappa_{AC}q^{eC}}{T} \,,
	\end{equation}
	where the latter equation is the multi-component form of the Wiedemann-Franz law.
	
	We note that Eq.\ \eqref{eq:cattaneo} is the simplest, and not the most general, equation of motion compatible with Eq.\ \eqref{TdS2}. In fact, adding some vectors $\mcM^A_\mu$ to either side of Eq.\ \eqref{eq:cattaneo}, where $ V_A^\mu \mcM^A_\mu  = 0$, does not affect Eq.\ \eqref{TdS2}. 
	Of course, one should keep in mind that the added term must vanish in equilibrium and be orthogonal to $u_\mu$.
	One possible term of such form is $\Omega_{\mu\nu}V_{C}^\nu b^{AC}$, where $\Omega_{\alpha \beta} \equiv \nabla_{[\alpha} u_{\beta]}$ is the fluid kinematic vorticity tensor. Adding this term to the right-hand side of Eq.\ \eqref{eq:cattaneo} gives rise to the so-called Coriolis term which is usually found in the literature \cite{Fotakis:2022usk}. Another possible term is proportional to $F_{\mu\nu}V_{C}^\nu b^{AC}$. 
	In a conducting fluid, such a term gives rise to a term proportional to $\epsilon^{\mu\nu\a\b}\mcB_\a V_{C\b}b^{AC}$ added to Eq.\ \eqref{eq:cattaneo}, where $\mcB^\mu = \tfrac{1}{2}\epsilon^{\mu\nu\a\b}u_\nu F_{\a\b}$ is the magnetic four-vector. Such a term, in the single component limit, can be found for example in Ref.\ \cite{Denicol:2019iyh}. 
	These terms are consistent with the second law of thermodynamics, but their nature and existence cannot be deduced from it.
	Furthermore, they do not modify the information current of the theory (which depends only on the constitutive relations \eqref{eq:const-rels}) and, therefore, do not play any role in the thermodynamic stability analysis discussed in the main text.
	
	For convenience, here we rewrite the stability condition of the single component diffusive Israel-Stewart theory (Eq. \eqref{eq:single-is-cc}) in terms of relaxation time $\tau$ and electric conductivity $\sigma$,
	\begin{eqnarray}
		{\tau} &>&  \frac{h\sigma}{q n^2}\left[\frac{1}{1-c_s^2}\left(1-\dfrac{n}{T}\pdv[2]{\ed}{n}{s}-\dfrac{s}{T}\pdv[2]{\ed}{s} \right)^2 + \dfrac{h}{T^2} \pdv[2]{\ed}{s} - 1\right] \,.
	\end{eqnarray}
	
	\section{Recovering Olson's information current}\label{app:olson-ic}

	Calculating $\d {T}^{\mu\nu}= \dot{T}^{\mu\nu}(0)$ for the stress-energy tensor of Eq.\ \eqref{eq:const-em-tensor}, contracting the result with $\d u_\nu$, and recalling that $u_\nu \d u^\nu  = 0$, we find
	\begin{equation}
		(\ed+P)u^\mu\delta u^\nu \delta u_\nu =
		\delta T\indices{^\mu _\nu}\delta u^\nu - \delta P \delta u^\mu  \, . 
	\end{equation}
	Inserting this into Eq.\ \eqref{eq:is-info-current} for $l=1$, we find 
	\begin{eqnarray}
		TE^\mu &=& \left[ \d T \d s +\d \mu \d n +\left(\ed+P\right)\d u^\nu \d u_\nu + b \, \d V^\nu \d V_{\nu} \right]\dfrac{u^\mu}{2} + \delta P\delta u^\mu +T\delta \left(\dfrac{\mu}{T} \right)\delta V^\mu \\\nonumber
		&=& \delta T\indices{^\mu _\nu}\delta u^\nu -\dfrac{1}{2}(\ed+P)u^\mu \delta u^\nu \delta u_\nu +T\delta \bigg(\dfrac{\mu}{T} \bigg)\delta V^\mu+\big[ \delta T \delta s +\delta \mu \delta n\big]\dfrac{u^\mu}{2}+\dfrac{1}{2}  b \, \delta V^\nu \delta V_{\nu} \, u^\mu \,. 
	\end{eqnarray}
	This information current coincides with the one of Olson \cite{OlsonLifsh1990} if one considers that
	\begin{equation}
		\delta T \delta s +\delta \mu \delta n =\dfrac{\partial \ed}{\partial P}\bigg|_{\mathfrak{s}} \dfrac{(\delta P)^2}{\ed+P} +\dfrac{\partial \ed}{\partial \mathfrak{s}}\bigg|_{P} \dfrac{\partial P}{\partial \mathfrak{s}}\bigg|_{\mu/T} \dfrac{(\delta \mathfrak{s})^2}{\ed+P} \, ,
	\end{equation}
	where $\mathfrak{s}=s/n$ is the specific entropy. 
	For proof of this identity, see the supplementary Material of Ref.\ \cite{GavassinoGibbs2021}.
	

	\section{Stability conditions from Carter-type information currents}\label{app:carter-ic}
	
	Consider a generic information current of the form ($X$ and $Y$ are species indices)
	\begin{equation}
		TE^\mu = \left[\ed^{XY} \delta n_X \delta n_Y + \mathcal{K}^{XY}\delta j_X^\nu \delta j_{Y\nu}  \right] \dfrac{u^\mu}{2} + \ed^{XY} \delta j_X^\mu \delta n_Y \, ,
	\end{equation}
	where the perturbations fields $\delta \varphi_i =\{\delta n_X, \delta j_X^\nu \}$ are a set of density-flux couples  (with $u_\mu \delta j_X^\mu=0$), while $\ed^{XY}$ and $\mathcal{K}^{XY}$ are symmetric background matrices. 
	We require $E^\mu$ to be future-directed timelike. 
	The isotropy of the problem implies that, if we work in the equilibrium local rest frame, we only need to demand $e=2T(E^0-E^1)$ to be positive for all non-vanishing perturbations. Explicitly,
	\begin{eqnarray}
		e&=& \ed^{XY} \delta n_X \delta n_Y + \mathcal{K}^{XY}\delta j_X^1 \delta j_Y^1-2\ed^{XY}\delta j_X^1 \delta n_Y+\mathcal{K}^{XY}\delta j_X^2 \delta j_Y^2+\mathcal{K}^{XY}\delta j_X^3 \delta j_Y^3 \nonumber\\
		&=& \ed^{XY}(\delta n_X{-}\delta j_X^1)(\delta n_Y{-}\delta j_Y^1)+(\mathcal{K}^{XY}{-}\ed^{XY})\delta j_X^1 \delta j_Y^1+ \mathcal{K}^{XY}\delta j_X^2 \delta j_Y^2+\mathcal{K}^{XY}\delta j_X^3 \delta j_Y^3 \, .
	\end{eqnarray}
	Clearly, $e$ is positive definite if and only if the symmetric matrices $\ed^{XY}$, $\mathcal{K}^{XY}$, and $\mathcal{K}^{XY}{-}\ed^{XY}$ are positive definite.
	
	\section{Details of derivation of Eq.\ \eqref{eq:mirable}}
	\label{app:mirable}
	
	To derive equation \eqref{eq:mirable-dot}, we study the three terms on the right-hand side of \eqref{eq:em-phidot-step-0} one by one. The trick is to express some selected Faraday tensors in terms of their vector potentials and to rearrange the derivatives using the Leibniz rule at the expense of collecting some perfect divergences. In the first term, we substitute the third factor using $F^{\nu \alpha}=\nabla^\nu A^\alpha-\nabla^\alpha A^\nu$, and invoke the Leibniz rule for $\nabla^\alpha$ to get
	\begin{eqnarray}
		\beta^\star_\nu \dot{F}^{\mu}_{\,\alpha}F^{\nu\alpha} =\dot{F}^{\mu \alpha} (\mathcal{L}_{\beta^\star} A)_\alpha +A^\nu \beta^\star_\nu \dot{J}^{e\mu} - \nabla_\alpha (\beta^\star_\nu A^\nu\dot{F}^{\mu \alpha}) \, .
	\end{eqnarray}
	In the second term, we substitute the third factor using $\dot{F}^{\nu \alpha}=\nabla^\nu \dot{A}^\alpha-\nabla^\alpha \dot{A}^\nu$, and invoke the Leibniz rule for both $\nabla^\alpha$ and $\nabla^\nu$ to obtain
	\begin{equation}
		\beta^\star_\nu F^{\mu}_{\,\alpha}\dot{F}^{\nu\alpha} =\dot{A}_\nu \beta^{\star \nu} J^{e\mu}-\dot{A}_\alpha (\beta^{\star \nu}\nabla_\nu F^{\mu \alpha}-F^{\mu \nu}\nabla_\nu \beta^{\star\alpha}) + \nabla_\alpha(\beta^{\star \alpha}F^{\mu \nu}\dot{A}_\nu -\dot{A}^\nu \beta^\star_\nu F^{\mu \alpha}) \, .
	\end{equation}
	Finally, we rewrite the third term in the form $-F^{\alpha \beta}(\nabla_\alpha \dot{A}_\beta)\beta^{\star \mu}$, and use the Leibniz rule for $\nabla_\alpha$ to get
	\begin{equation}
		- \frac{1}{2}F^{\alpha \beta}\dot{F}_{\alpha \beta} \beta^{\star \mu} = -\dot{A}_\nu J^{e\nu} \beta^{\star \mu} + \dot{A}_\alpha F^{\nu \alpha}\nabla_\nu \beta^{\star \mu} - \nabla_\alpha(\beta^{\star \mu} F^{\alpha \nu} \dot{A}_\nu) \, .
	\end{equation}
	Adding up all three pieces, we finally recover equation \eqref{eq:mirable-dot}, with
	\begin{eqnarray}\label{eq:z1}
		Z_{(1)}^{[\alpha \mu]}= -\beta^\star_\nu A^\nu \dot{F}^{\mu \alpha} - \beta^\star_\nu \dot{A}^\nu F^{\mu \alpha} + 2\beta^{\star [\alpha}F^{\mu]\nu} \dot{A}_\nu \, .
	\end{eqnarray}
	The derivation of \eqref{eq:mirable-ddot} is completely analogous to the above. 
	The corresponding residual terms are given by
	\begin{eqnarray}\label{eq:z2}
		Z_{(2)}^{[\alpha \mu]}= -\beta^\star_\nu A^\nu \ddot{F}^{\mu \alpha} - \beta^\star_\nu \ddot{A}^\nu F^{\mu \alpha} + 2\beta^{\star [\alpha}F^{\mu]\nu} \ddot{A}_\nu \, .
	\end{eqnarray}
	Since $Z_{(1)}^{[\alpha \mu]}$ and $Z_{(2)}^{[\alpha \mu]}$ are antisymmetric, Gauss' theorem applies, see \citet{Poisson_notes}, \S 3.3.3, equation (3.3.3).
	\label{lastpage}

	\bibliography{main}
\end{document}

%% file: commands.tex
\def\ed{\varepsilon}

\def\diag{\mathrm{diag}}

\newcommand{\mcB}{{\mathcal{B}}}

\newcommand{\mcE}{{\mathcal{E}}}

\newcommand{\mcK}{{\mathcal{K}}}
\newcommand{\mcL}{{\mathcal{L}}}
\newcommand{\mcM}{{\mathcal{M}}}

\newcommand{\mcR}{{\mathcal{R}}}

\def\a{\alpha}
\def\b{\beta}

\def\d{\delta}

\def\mnras{{MNRAS}}             
\def\prd{{Phys.~Rev.~D}}        
\def\prl{{Phys.~Rev.~Lett.}}    
\newcommand{\lieder}[2]{\mcL_{#1}{#2}}

\def\a{\alpha}
\def\b{\beta}

\def\d{\delta}

\def\rabbit{\mcR}
\def\dermat{\varepsilon}